\documentclass[epjST]{svjour}
\usepackage{graphicx}
\usepackage{dcolumn}
\usepackage{caption}
\usepackage{array}
\usepackage{tabularx}
\usepackage[numbers,sort&compress]{natbib}

\usepackage{floatrow}
\usepackage[labelsep=space]{caption}

\usepackage{eucal}
\usepackage[dvips]{epsfig}
\usepackage{amssymb}

\usepackage{amsmath,bm}
\newcommand{\be}{\begin{eqnarray}}
\newcommand{\ee}{\end{eqnarray}}
\newcommand{\bse}{\begin{subequations}}
\newcommand{\ese}{\end{subequations}}

\newcommand{\bnum}{\begin{enumerate}}
\newcommand{\enum}{\end{enumerate}}

\newcommand{\bit}{\begin{itemize}}
\newcommand{\eit}{\end{itemize}}

\newcommand{\bc}{\begin{cases}}
\newcommand{\ec}{\end{cases}}

\newcommand{\bpm}{\begin{pmatrix}}
\newcommand{\epm}{\end{pmatrix}}

\newcommand{\bvm}{\begin{vmatrix}}
\newcommand{\evm}{\end{vmatrix}}

\newcommand{\bs}{\boldsymbol}

\newcommand{\gb}{\beta}
\newcommand{\gc}{\gamma}

\newcommand{\gl}{\lambda}

\newcommand{\gs}{\sigma}

\newcommand{\Gc}{\Gamma}

\newcommand{\Gs}{\Sigma}
\newcommand{\Gl}{\Lambda}

\newcommand{\p}{\partial}
\newcommand{\f}{\frac}

\usepackage{color}

\begin{document}
\title{Generalized Navier-Stokes equations for active suspensions}
\author{Jonasz S\l{}omka\inst{1}\fnmsep\thanks{\email{jslomka@mit.edu}} \and J\"orn Dunkel\inst{1}\fnmsep\thanks{\email{dunkel@mit.edu}}}
\institute{Department of Mathematics, Massachusetts Institute of Technology,\\
77 Massachusetts Avenue E17, Cambridge, MA 02139-4307, USA 
}
\abstract{
We discuss a minimal generalization of the incompressible  Navier-Stokes equations to describe the solvent flow in an active suspension. To account phenomenologically for the presence of an active component driving the ambient fluid flow, we postulate 
a generic nonlocal extension of the stress-tensor, conceptually similar to those recently introduced in granular media flows. 
Stability and spectral properties of the resulting hydrodynamic model are studied both analytically and numerically for the two-dimensional~(2D) case with periodic boundary conditions. Future generalizations of this momentum-conserving theory could be useful for quantifying the shear properties of active suspensions.} 
\maketitle
%
\section{Introduction}
\label{intro}

An active suspension~\cite{2013Marchetti_Review,2007Cisneros,Swinney_bactclust,2010Bausch,2012Sokolov,2012Sanchez_Nature,2013Dunkel_PRL} is, roughly speaking,  a passive fluid medium that contains at least one \lq micro-swimmer\rq{} species capable of converting chemical into kinetic energy.  If the swimmer concentration is sufficiently high, their collective dynamics can induce rich non-equilibrium flow patterns in the ambient fluid~\cite{2004DoEtAl,2008Wolgemuth,2009BaMa_PNAS,2010SaintillanShelley,2012Wensink,2012Woodhouse,2013Lauga,2013Giomi_PRL,2014Zoettl_PRL}, thereby causing significant changes in the transport properties~\cite{2000WuLi,2009LeEtAl_Gold,2010DuPuZaYe,2011ZaDuYe} and rheological response~\cite{2009SoAr,2009Rafai,2011Aranson_PRE,2012Foffano_EPJE} of the solvent medium. An intriguing, seemingly generic feature of dense active suspensions is the emergence of a characteristic topological defect or vortex distance~\cite{2012Sokolov,2012Sanchez_Nature,2013Dunkel_PRL,2012Wensink,2013Wioland_PRL,2013Thampi_PRL},  thought to arise from the competition between self-propulsion, steric and hydrodynamic interactions~\cite{2014Lushi_PNAS}. Although the microscopic origins of such dynamical length-scale selection mechanisms are not yet fully understood, their experimentally confirmed presence~\cite{2012Sokolov,2012Sanchez_Nature,2013Dunkel_PRL,2012Wensink,2013Wioland_PRL} suggests  that one can effectively describe active suspensions in terms of  \lq non-local\rq{} higher-than-second-order partial differential equations (PDEs)~\cite{2013Dunkel_NJP}, in analogy with well-established continuum theories of pattern formation in elastic materials~\cite{Landau59}, granular media~\cite{2006Aranson} and convection phenomena~\cite{1977SwiftHohenberg}.

\par
Indeed, recent experimental and theoretical studies~\cite{2013Dunkel_PRL,2012Wensink} confirm that a fourth-order extension of the Toner-Tu theory~\cite{1995TonerTu_PRL,1998TonerTu_PRE} can reproduce, both qualitatively and quantitatively, many of the main statistical features of dense bacterial suspensions. 
Specifically, these experiments~\cite{2013Dunkel_PRL,2012Wensink}  measured the mean \emph{bacterial} velocity field $\bs u(t,\bs x)$, which can be approximately decomposed in the form $\bs u(t,\bs x)\simeq \bs v(t,\bs x)+v_0\bs P(t,\bs x)$, where $\bs v(t,\bs x)$ is the underlying \emph{solvent} velocity field and $\bs P(t,\bs x)$ denotes the local mean orientation of the bacteria. The parameter $v_0$ is the typical bacterial self-swimming speed \emph{relative to the solvent flow} (in general, $v_0$ is also a fluctuating quantity).  The  bacterial velocity data $\{\bs u\}$, obtained by standard PIV methods~\cite{2004DoEtAl,2007Cisneros,2012Sokolov}, were found to agree well with predictions of the incompressible  fourth-order theory~\cite{2013Dunkel_PRL,2012Wensink} 
\bse\label{e:bac_flow}
\be
\label{e:bac_flow-a}
\nabla\cdot \bs u &=&0 \\
 (\partial_t + \lambda_0 \boldsymbol u\cdot \nabla) \boldsymbol u
 &=&
-\nabla (p  -\lambda_1 \boldsymbol u^2) - \gb (\boldsymbol u^2-u_0^2)\boldsymbol u + 
\Gamma_0 \nabla^2 \boldsymbol u -\Gamma_2(\nabla^2)^2    \boldsymbol u,
\ee\label{e:conti-model}
\ese
where the pressure $p(t,\bs x)$ is the Lagrange multiplier for the incompressibility (bacterial mass conservation) constraint~\eqref{e:bac_flow-a}.
The parameter $\gl_0$ describes nematic advection and $\gl_1$ an active pressure contribution. The $(\beta, u_0)$-terms correspond to a quartic Landau-type velocity potential~\cite{1998TonerTu_PRE,2005ToTuRa,2010Ramaswamy} and account for the formation of locally aligned  bacterial jets~\cite{2007Cisneros}.  The nonlocal $(\Gc_0,\Gc_2)$-terms encode passive and active stresses due to hydrodynamic and steric interactions, and determine the characteristic vortex size $\Gl_\Gc\simeq 2\pi \sqrt{\Gc_2/(-\Gc_0)}$  in  the model when $\Gc_2>0$ and $\Gc_0<0$.  Conceptually, Eqs.~\eqref{e:conti-model} extend the incompressible Toner-Tu theory~\cite{1998TonerTu_PRE,2005ToTuRa,2010Ramaswamy} through the additional Swift-Hohenberg-type~\cite{1977SwiftHohenberg} instability that arises for $\Gc_0<0$. A similar linear instability mechanism was recently derived by Gro\ss{}mann \textit{et al.}~\cite{2014Grossmann_PRL} who considered a self-propelled particle model with velocity-dependent interaction. With regard to our subsequent discussion, it is important to note that Eqs.~\eqref{e:bac_flow}, which are defined in the rest-frame of the microfluidic channel confining the suspensions,  are non-conservative due to the aligning $\beta$-term, reflecting the fact that the self-swimming field $v_0\bs P$ is \emph{not} a conserved quantity (just as in the classical Toner-Tu model~\cite{1995TonerTu_PRL,1998TonerTu_PRE}).

\par
Although Eqs.~\eqref{e:bac_flow} give satisfactory predictions for the \emph{bacterial} velocity field  $\bs u(t,\bs x)$ in a stationary setting~\cite{2013Dunkel_PRL,2012Wensink}, they are of limited use with regard to shear experiments, which typically measure the response of the \emph{solvent} flow in the presence of moving boundaries.  Aiming to develop a simplified phenomenological framework for the future mathematical description of rheological measurements~\cite{2009SoAr,2009Rafai}, we will focus here on the complementary problem of constructing effective models for the solvent velocity field~$\bs v(t,\bs x)$.  Specifically, we are interested in identifying a minimal extension of the Navier-Stokes (NS) equations that reproduces  qualitatively the experimentally observed, turbulent tracer dynamics in active suspensions~\cite{2013Dunkel_PRL}. In contrast to traditional approaches that build on  explicit couplings between solvent flow and additional orientational order-parameter fields~\cite{2008Wolgemuth,2009BaMa_PNAS,2010SaintillanShelley,2013Giomi_PRL,2012Foffano_EPJE,2002Ra}, we investigate here analytically and numerically higher-order \textit{ad hoc} closure conditions for the stress tensor.  Despite some technical differences, our approach shares conceptual similarities with the recently proposed,  effectively non-local constitutive relations that have led to promising progress in the quantitative understanding of porous media flows~\cite{2013Henann_PNAS}.

\section{Generalized Navier-Stokes (NS) model}
\label{sec:1}
We focus on a coarse-grained model of active micro-swimmer suspensions, assuming that a single velocity field, $\bm v(t,\bs x)$, describes the solvent flow on scales several times larger than an individual micro-swimmer. Considering incompressible solvent  flow, we assume that the dynamics of  $\bm v(t,\bs x)$ is governed by the mass and momentum conservation laws
\bse\label{e:eom1}
\be
0&=&\nabla\cdot \bs v,\\
\p_t \bs v+ (\bs v\cdot\nabla)\bs v
&=& 
-\nabla p  +\nabla\cdot\bs \gs,
\ee
\ese
with scalar pressure $p(t,\bs x)$  and symmetric stress tensor $\bs \gs(t,\bs x)$. As usual,  the swimmers are assumed to drive solvent flow by modifying the stress field $\bs \gs$.  However, instead of constructing $\bs \gs$ from  orientational order-parameter fields~\cite{2008Wolgemuth,2009BaMa_PNAS,2010SaintillanShelley,2013Giomi_PRL,2012Foffano_EPJE,2002Ra}, we hypothesize that the stress generated by the micro-swimmers can be captured through a generic closed-form ansatz
\be
\bs\sigma=\bs \Gs(\nabla,\bs v).
\ee
In this paper, we will focus on a representative of the class of isotropic traceless tensors\footnote{More generally, 
one could also consider additional quasi-nematic stress contributions~\mbox{$\propto \bs v\bs v-\bs I |\bs v|^2/d$}, where  $\bs I$ is the $d$-dimensional identity tensor. Such terms would effectively rescale the advective derivative and add a kinetic pressure contribution.}
\be\label{e:f_ansatz}
\bs \Gs(\nabla,\bs v)=f(\nabla^2)\left[ (\nabla \bs v)+(\nabla\bs v)^\top\right],
\ee
where $\nabla^2$ is the Laplace operator, and the scalar function $f(\cdot)$  quantifies the swimmers-solvent coupling.   Intuitively, Eq.~\eqref{e:f_ansatz} can be thought to arise from a truncated gradient expansion of an integral kernel representation of  the \lq full\rq{} stress tensor $\bs \gs$, similar to a Kramers-Moyal expansion~\cite{1996Risken}.  In the limit case of a constant function $f(\nabla^2)\equiv \Gamma_0$, corresponding to a passive isotropic fluid,  Eqs.~(\ref{e:eom1}) reduce to the standard Navier-Stokes equations. 

\par
In the remainder, we will  restrict the discussion to symmetric second-order polynomials
\be\label{e:f_poly_ansatz}
f(\nabla^2)&=&\Gamma_0-\Gamma_2(\nabla^2)+\Gamma_4(\nabla^2)^2.
\ee
The constants  $\Gamma_0$ and $\Gamma_4$ are assumed to be positive to ensure asymptotic stability, whereas the parameter $\Gamma_2$ may have either sign.  Nontrivial steady-state flow structures emerge for negative values $\Gamma_2<0$. Inserting Eqs.~\eqref{e:f_ansatz} and \eqref{e:f_poly_ansatz} into Eqs.~\eqref{e:eom1}, we obtain the hydrodynamic equations 
\bse
\label{e:eom2}
\be
\label{e:eom2a}
0&=&\nabla\cdot \bs v,
\\
\label{e:eom2b}
\p_t \bs v+ (\bs v\cdot\nabla)\bs v
&=& 
-\nabla p  +\Gamma_0 \nabla^2 \bs v-\Gamma_2 \nabla^4 \bs v+\Gamma_4 \nabla^6 \bs v,
\ee
\ese
where $\nabla^{2n}\equiv(\nabla^2)^{n}$ from now on. Below, we analyze the generalized NS equations~\eqref{e:eom2} on a square domain. Since we aim to understand their bulk behavior, we adopt periodic boundary conditions throughout.

\section{Analytical results}
\label{sec:BasicInsights}

To obtain some intuition about the generalized NS model~\eqref{e:eom2}, we first note that its linear part supports a stationary vortex lattice of period \mbox{$\sim\sqrt{\Gamma_4/(-\Gamma_2)}$}, when $\Gamma_2$ is negative. This follows from the fact that
\be
\Big(\Gamma_0 \nabla^2 -\Gamma_2 \nabla^4 +\Gamma_4 \nabla^6 \Big) e^{i \bs k \cdot \bs x}=0
\ee
if $k=|\bs k|$ is one of the roots of $\Gamma_0 +\Gamma_2 k^2 +\Gamma_4 k^4=0$, with real positive roots existing only when $\Gamma_2<0$.  Furthermore, since the nonlinear advective terms on the lhs. of Eq.~\eqref{e:eom2b} will generally lead to mixing, one can expect to find parameters such that Eqs.~\eqref{e:eom2}  produce turbulent mesoscale patterns similar to those observed  in experiments~\cite{2012Sokolov,2013Dunkel_PRL,2012Wensink}.

\par
To perform a more detailed stability analysis, we focus on a periodic square domain $\Omega=[-L/2,L/2]^2$ and rescale Eqs.~\eqref{e:eom2} by introducing dimensionless quantities
\be
\bs x\to \f{2\pi}{L}\bs x,\quad t\to \f{(2\pi)^2\Gamma_0}{L^2}t,\quad \bs v\to \f{L}{2\pi \Gamma_0}\bs v
, \quad p\to \f{L^2}{(2\pi)^2\Gamma_0^2}p,\quad \bs k\to \f{L}{2\pi}\bs k.
\ee
One then finds that the dynamics of the model is characterized by the two dimensionless groups
\be
\gamma=\f{\Gamma_0\Gamma_4}{\Gamma_2^2},
\qquad
\gamma_2=\f{(2\pi)^2\Gamma_2}{L^2\Gamma_0},
\ee
and the rescaled Eqs.~\eqref{e:eom2} take the form
\bse
\label{e:eomND}
\be
\nabla\cdot \bs v &=&0 \\
\p_t \bs v+ (\bs v\cdot\nabla)\bs v
&=& 
-\nabla p  +\nabla^2\bs v-\gamma_2\nabla^4\bs v+\gamma\gamma_2^2\nabla^6\bs v.
\label{e:eomND-b}
\ee
\ese
To ensure stability at short wavelengths, $\gamma$ must be always positive.  Nontrivial flow structures require $\gamma_2<0$. We expect that there is a region in the  $(\gamma,\gamma_2)$-parameter space that supports a quasi-chaotic steady-state dynamics characterized by the creation and annihilation of vortices.
\par
To estimate this parameter region, we investigate how the total kinetic energy, $E(t)=\f{1}{2}\int_\Omega d\bs x\, |\bs v|^2$, varies with time. From the equations of motion with periodic boundary conditions, one finds
\be
\dot{E}(t)&=&\notag
\int_\Omega d\bs x\, \bs v \cdot\p_t \bs v \\
&=&\notag
\int_\Omega d\bs x\, \left[-\nabla\cdot \left(\f{1}{2}|\bs v|^2 \bs v+p\bs v\right) + \bs v \cdot (\nabla^2\bs v-\gamma_2\nabla^4\bs v+\gamma\gamma_2^2\nabla^6\bs v)\right]
\\
&=&
\int_\Omega d\bs x \,\bs v \cdot \left( \nabla^2\bs v-\gamma_2\nabla^4\bs v+\gamma\gamma_2^2\nabla^6\bs v\right).
\ee
We next insert the Fourier series for the velocity, $\bs v(t,\bs x)=\mathcal{F}^{-1}(\hat{\bs v})\equiv\sum_{\bs k}e^{i \bs k\cdot \bs x}\hat{\bs v}(t, \bs k)$, where $\bs k \in \mathbb{Z}^2$, to obtain
\be
\dot{E}(t)
=
-(2\pi)^2\sum_{\bs k} k^2\big(1+ \gamma_2 k^2  +\gamma\gamma_2^2 k^4\big)|\hat{v}(t,\bs k)|^2.
\ee
The first and the third term in the brackets are always positive, and therefore dissipate energy. If $\gamma_2>0$ holds, then $\dot{E}<0$ always; in this case, any solvent flow in the system is rapidly damped out. More interestingly, however, when $\gamma_2<0$,  the active component pumps energy into the flow. We may then ask if, at least for some region in the $(\gamma,\gamma_2)$-plane, the energy input and the energy dissipation can balance each other. If such a steady-state exists, then the total system energy should fluctuate about a constant mean value. More formally, we expect in this case that the mean energy change vanishes,
\bse
\be
\langle \dot{E} \rangle_{\Delta,T}\equiv\lim_{\Delta\to\infty}\lim_{T\to\infty} \f{1}{\Delta}\int_{T}^{T+\Delta} \dot{E}(t)dt
\; \to \;0,
\ee
and that the time-averaged Fourier coefficients become stationary and isotropic,
\be
\langle |\hat{\bs v}(t,\bs k)|^2\rangle_{\Delta,T} \;\to\; \langle |\hat{\bs v}(k)|^2\rangle.
\ee
\ese
Provided that the steady-state is attainable, we have the following energy balance equation
\bse
\be\label{e:EnergyBalance}
\sum_{k} k^2\big(1+ \gamma_2 k^2  +\gamma\gamma_2^2 k^4\big)\mathcal{E}(k)=0,
\ee
where the energy spectrum, $\mathcal{E}(k)$, is defined as~\cite{2004Frisch}
\be
\left\langle\f{1}{2}\int_\Omega d\bs x\, |\bs v|^2\right\rangle_{\Delta,T}=\sum_{k}\mathcal{E}(k),
\ee
\ese
yielding 
\be
\mathcal{E}(k)=\f{(2\pi)^2}{2}\sum_{\bs k':|\bs k'| =k}\langle|\hat{\bs v}(k')|^2\rangle.
\ee

\par
Building on the above considerations, we can now analytically estimate the part of the $(\gamma,\gamma_2)$-parameter plane where a steady-state can be reached. Neglecting pressure, the linearized version of Eq.~\eqref{e:eomND-b} reads
\be
\p_t \bs v&=& 
\nabla^2\bs v-\gamma_2\nabla^4\bs v+\gamma\gamma_2^2\nabla^6\bs v.
\ee
Using the Fourier series for $\bs v$ as before, this is equivalent to
\bse
\be
\p_t \hat{\bs v}(t, \bs k)&=&\Lambda(k)\,\hat{\bs v}(t, \bs k), 
\ee
where
\be
\Lambda(k)&=&-k^2(
1+\gamma_2 k^2+\gamma\gamma_2^2k^4)
\notag\\
&=&
-\gamma \gamma_2^2 k^2(k^2-k_-^2)(k^2-k_+^2).
\ee
\ese
The range of physically reasonable zeros $k_\pm$ can be inferred from stability considerations as follows:
\par
Fourier modes may become unstable if the micro-swimmers inject a sufficiently large amount of energy into the system. In the linearized model, such supercritical energy injection leads to divergence at an exponential rate. However, the advective term in the full nonlinear system mixes different modes, facilitating energy dissipation through the decaying modes. The stability of a given mode is determined by the sign of $\Lambda(k)$. To observe non-trivial flow structures, we require $\Lambda(k)>0$ for some $k>0$, implying that \mbox{$\gamma_2<0$} and $\gamma<1/4$, because otherwise $(1+ \gamma_2 k^2  +\gamma\gamma_2^2 k^4)>0$ 
and, hence,  no real positive roots~$k_{\pm}$ exist. Furthermore, its is plausible to assume that any realistic active system has a long-wavelength cut-off corresponding to the largest scale at which energy is collectively injected into the fluid,  implying  that the system should be dissipative in some vicinity of its lowest mode, $k=1$. Otherwise, there is no room left for an inverse dissipative energy cascade and the energy could continuously accumulate at long wavelengths. We therefore demand $k_->1$, which implies that $\gamma_2>\f{1}{-2\gamma}(1-\sqrt{1-4\gamma})$. Thus, all discrete  $\bs k$-modes lying in the annular region $1<k_-<k<k_+$ inject energy, whereas the complementary set of modes dissipates energy. For instance, if we adopt the simplifying assumption that, as in classical 2D turbulent flow, $\mathcal{E}(k)$ scales as $k^{-5/3}$ up to some cut-off mode, which we take to be the largest unstable mode, $k^2_{+}=\f{1}{-2\gamma\gamma_2}(1+\sqrt{1-4\gamma})$, then, after approximating the sum by an integral, Eq.~\eqref{e:EnergyBalance} predicts the critical value of $\gamma=0.16$. \par
\begin{figure*}
\resizebox{0.98\textwidth}{!}{
  \includegraphics{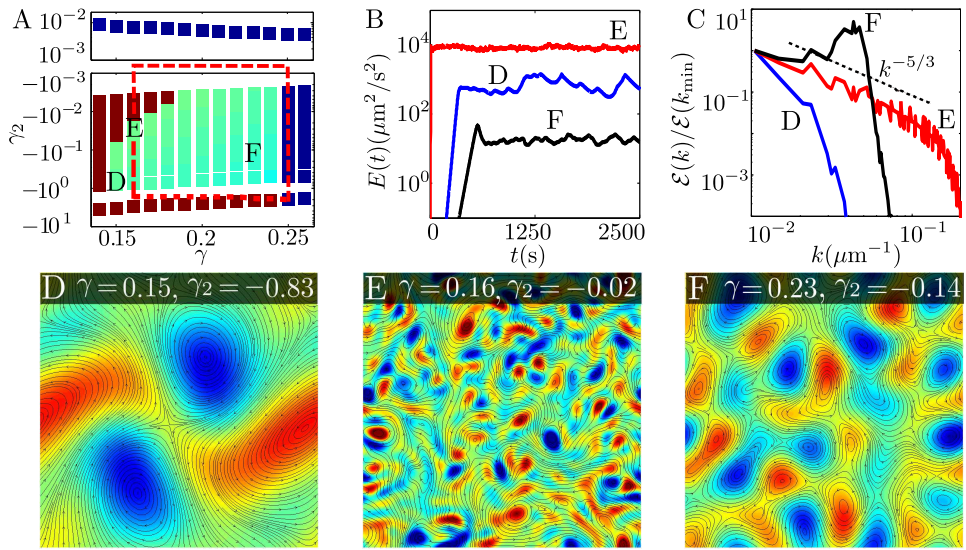}
}
\caption{Simulation results for the rescaled generalized Navier-Stokes model defined in Eqs.~\eqref{e:eomND}. A: Numerical stability analysis in the $(\gc,\gc_2)$-parameter plane. Nontrivial energetically stable steady-state solutions exist in the green region (labels D, E and F correspond to snapshots shown in panels D-F). The numerical results agree well with the analytically predicted region of flow structure formation (red dashed lines) obtained in Sec.~\ref{sec:BasicInsights}. In the purely dissipative regime (dark blue), the system converges to the zero-flow solution. The truncated model becomes unstable (brown region) when the dissipation is not able to balance the energy input.   B, C: Kinetic energy as a function of time, $E(t)$, and corresponding normalized energy spectra, $\mathcal{E}(k)$, for the three parameter choices in the bottom row. D-F: Snapshots of the steady-state flow stream lines for simulations with periodic boundary conditions. The background color represents the associated vorticity fields $\omega = \nabla \wedge \bs v$, normalized by their  maximal values. The domain size is $L\times L$, with $L=600\,\mu\text{m}$ in physical units.}
\label{fig:landscape}
\end{figure*}

In summary, these considerations suggest that an energetically stable, nontrivial  steady-state can be reached in the parameter range  $0.16<\gamma<0.25$ and  $0>\gc_2>\f{1}{-2\gamma}(1-\sqrt{1-4\gamma})$. These simple estimates agree well with our numerical results, as shown in Fig.~\ref{fig:landscape}A.

\section{Stream function formulation and numerical implementation}

To solve Eqs.~\eqref{e:eomND} numerically, it is convenient to reformulate the dynamical equations~\eqref{e:eomND} in terms of a stream function. By means of the Helmholtz-Hodge decomposition \cite{Abraham1988}, we can express the solvent flow field $\bs v$ as a sum of divergence-free, curl-free and harmonic components,
\be\label{e:flowdecomposition}
\bs v=\nabla\Phi+\nabla\wedge\Psi+\bs V,
\ee
for some scalar functions $\Phi(t,\bs x)$ and $\Psi(t,\bs x)$ and a harmonic vector field $\bs V(t,\bs x)$ satisfying $\nabla^2 \bs V=0$. On periodic domains, harmonic functions are constant, so $\bs V$ is interpreted as the fluid center of mass velocity. We will always work in the center of mass frame, hence $\bs V=0$ from now on. Similarly, $\Phi$ is also a constant since incompressibility implies that $\nabla^2\Phi=0$, so that $\Phi$ is a harmonic function. Thus, Eq.~\eqref{e:flowdecomposition} reduces to $\bs v=\nabla\wedge \Psi$.

\par
To obtain the evolution equation for $\Psi$, we take the divergence and the curl of Eq.~\eqref{e:eomND-b}, which gives
\bse
\be
\label{e:streamfn1}
(\nabla\nabla\Psi) :(\nabla\nabla\Psi) -(\nabla^2 \Psi)^2&=&-\nabla^2 p,
\\
\label{e:streamfn2}
\p_t(\nabla^2 \Psi)+\nabla(\nabla^2\Psi)\wedge \nabla\Psi&=&\nabla^4 \Psi -\gamma_2\nabla^6 \Psi+\gamma\gamma_2^2\nabla^8 \Psi.
\ee
\ese
The main advantage of this reformulation lies in the fact that the stream function~$\Psi$ is now the only dynamical variable, as the pressure $p$ can always be recovered from~$\Psi$ by solving the Poisson equation~\eqref{e:streamfn1}.

\par
The governing equation for the stream function, Eq.~\eqref{e:streamfn2}, can be solved with standard spectral methods~\cite{Canuto1987}. Using the Fourier series representation, the modes of~$\Psi$ evolve according to
\bse\label{e:FTPsi}
\be
\p_t \hat{\Psi}
+\mathcal{N}
&=&
-k^2(1+\gamma k^2+\gamma \gamma_2^2 k^4)\hat{\Psi}, 
\ee
where the nonlinear terms are abbreviated by
\be
\mathcal{N}&=&k^{-2}\mathcal{F}\{\mathcal{F}^{-1}(i\bs k k^2\hat{\Psi})\wedge
\mathcal{F}^{-1}(i\bs k\hat{\Psi})\}.
\ee
\ese
We integrated Eqs.~\eqref{e:FTPsi} with a classical fourth-order Runge-Kutta scheme, and approximated $\mathcal{F}(\cdot)$ by a Discrete Fourier Transform (DFT). The nonlinear term $\mathcal{N}$ is evaluated by inverting the DFT, performing the multiplication in position space, and then applying the DFT again. The aliasing error generated in this procedure is removed at the expense of using a larger number of Fourier modes and zero-padding when necessary (see~\cite{Canuto1987} for a detailed description of this method; in our simulations, we used a symmetric grid of $243\times 243$ modes).

\section{Results}

We first investigated the stability of numerical solutions of Eqs.~\eqref{e:eomND} on a periodic $L\times L$ square domain by performing systematic scan of the $(\gamma,\gamma_2)$-parameter plane,  initializing each simulation run with small random stream function values.  In agreement with our analytical considerations  in Sec.~\ref{sec:BasicInsights}, we found three qualitatively different asymptotic behaviors: 
\par
(i) For $\gamma_2>0$ or when $\gc$ becomes too large, the solvent dynamics is purely dissipative and approaches a stationary state of vanishing flow $\bs v\equiv 0$  (blue symbols in Fig.~\ref{fig:landscape}A).
\par
(ii)  For $\gamma_2<0$ and $\gamma$ too small, the solutions become unstable, reflected by an exponential blow-up of the kinetic energy  (brown symbols in Fig.~\ref{fig:landscape}A). 
\par
(iii) For $\gamma_2<0$ and moderate values of $\gamma$ the system exhibits quasi-chaotic steady-state flow patterns (green symbols in Fig.~\ref{fig:landscape}A). The numerically estimated boundaries of this physically relevant domain agree well with the analytical estimates (red dashed lines) from  Sec.~\ref{sec:BasicInsights}.
\par

Example snapshots of quantitatively different flow patterns for three parameter pairs $(\gamma,\gamma_2)$ are shown in Fig.~\ref{fig:landscape}D-F.  For all three parameter pairs, the kinetic energy approaches a constant mean value  (Fig.~\ref{fig:landscape}B). As $\gamma$ increases, the corresponding energy spectra develop  peaks near the linearly most unstable wave number, $k_* \sim -(2\gamma\gamma_2)^{-1}$ (black curve F in Fig.~\ref{fig:landscape}C). In position space, these peaks   correspond to a characteristic vortex size $\sim k_*^{-1}$  (Fig.~\ref{fig:landscape}F). Near $\gamma=0.16$, the spectra are well approximated by Kolmogorov scaling $\mathcal{E}(k)\propto k^{-5/3}$ with sharp cut-off (red curve E in Fig.~\ref{fig:landscape}C), in agreement with the assumptions made in Sec.~\ref{sec:BasicInsights} to derive the left vertical red dashed line in Fig.~\ref{fig:landscape}A.

\par
To relate the dimensionless parameters to physical relevant dimensional values, we may fix $\Gamma_0=10^{2}\,\mu\text{m}^2/\text{s}$ and identify the box length with $L=600\,\mu\text{m}$. With these choices, the typical steady-state speeds are in the range of $1\,\mu \text{m}/\text{s}$ to $100\, \mu \text{m}/\text{s}$,  as typical of passive tracer particles in dense bacterial suspensions~\cite{2013Dunkel_PRL}.
\par
Finally, we would still like to emphasize that the unstable regime (brown symbols in Fig.~\ref{fig:landscape}A) arises from the particular truncated polynomial ansatz in Eq.~\eqref{e:f_poly_ansatz}. Future quantitative comparison with experiments should focus on reconstructing better approximations of the function $f$ or, more generally, $\bs \gs$. Conversely, however, stability criteria provide useful physical constraints  for effective models that can be utilized in parameter estimation procedures.

\section{Summary}

We proposed and analyzed a minimal generalization of the Navier-Stokes equation to describe the solvent flow in an active suspension. The main assumption underlying this simple momentum-conserving model is that, on scales larger than the swimming cells or filaments, the complex fluid-swimmer interactions can be effectively captured by a generalized form of the stress energy tensor, which can be expanded in terms of higher-order differential operators. In this contribution, we focussed on a simple  example, corresponding to a sixth-order PDE, that produces turbulent flow features that are qualitatively similar to those observed through passive tracer-particle tracking in recent experiments~\cite{2013Dunkel_PRL}. 
\par
A future goal is to embed this class of models into a shear-flow setting as relevant to viscosity measurements~\cite{2009Rafai,2009SoAr}. This problem is conceptually and numerically nontrivial as appropriate boundary conditions need to be identified and implemented. Notwithstanding, such a phenomenological approach may help us to progress towards a sufficiently-general-yet-reasonably-simple framework for the classification of rheological observations in active suspensions. From a practical perspective, an interesting challenge will be to reconstruct an empirical form of the generalized stress-tensor $\bs \gs$ from experimentally measured solvent flow data.

\section*{Acknowledgements}
The authors would like to thank Anand Oza and Sebastian Heidenreich for advice on the numerical implementation. 
They are grateful to Igor Aranson, Markus B\"ar, Raymond Goldstein, Hartmut L\"owen, Lutz Schimansky-Geier, Holger Stark, Rik Wensink and  Julia Yeomans for helpful discussions.


\begin{thebibliography}{42}

\bibitem{2013Marchetti_Review}
M.C. Marchetti, J.F. Joanny, S.~Ramaswamy, T.B. Liverpool, J.~Prost, M.~Rao,
  R.A. Simha, Rev. Mod. Phys. \textbf{85}, 1143 (2013)

\bibitem{2007Cisneros}
L.H. Cisneros, R.~Cortez, C.~Dombrowski, R.E. Goldstein, J.O. Kessler, Exp.
  Fluids \textbf{43}, 737 (2007)

\bibitem{Swinney_bactclust}
H.P. Zhang, A.~Be'er, E.L. Florin, H.L. Swinney, Proc. Natl. Acad. Sci. USA
  \textbf{107}, 13626 (2010)

\bibitem{2010Bausch}
V.~Schaller, C.~Weber, C.~Semmrich, E.~Frey, A.R. Bausch, Nature \textbf{467},
  73 (2010)

\bibitem{2012Sokolov}
A.~Sokolov, I.S. Aranson, Phys. Rev. Lett. \textbf{109}, 248109 (2012)

\bibitem{2012Sanchez_Nature}
T.~Sanchez, D.T.N. Chen, S.J. DeCamp, M.~Heymann, Z.~Dogic, Nature
  \textbf{491}, 431 (2012)

\bibitem{2013Dunkel_PRL}
J.~Dunkel, S.~Heidenreich, K.~Drescher, H.H. Wensink, M.~B\"ar, R.E. Goldstein,
  Phys. Rev. Lett. \textbf{110}, 228102 (2013)

\bibitem{2004DoEtAl}
C.~Dombrowski, L.~Cisneros, S.~Chatkaew, R.E. Goldstein, J.O. Kessler, Phys.
  Rev. Lett. \textbf{93}(9), 098103 (2004)

\bibitem{2008Wolgemuth}
C.W. Wolgemuth, Biophys. J. \textbf{95}, 1564 (2008)

\bibitem{2009BaMa_PNAS}
A.~Baskaran, M.C. Marchetti, Proc. Natl. Acad. Sci. \textbf{106}(37), 15567
  (2009)

\bibitem{2010SaintillanShelley}
D.~Saintillan, M.~Shelley, J. R. Soc. Interface \textbf{9}(68), 571 (2011)

\bibitem{2012Wensink}
H.H. Wensink, J.~Dunkel, S.~Heidenreich, K.~Drescher, R.E. Goldstein,
  H.~L\"{o}wen, J.M. Yeomans, Proc. Natl. Acad. Sci. USA \textbf{109}(36),
  14308 (2012)

\bibitem{2012Woodhouse}
F.G. Woodhouse, R.E. Goldstein, Phys. Rev. Lett. \textbf{109}, 168105 (2012)

\bibitem{2013Lauga}
T.~Brotto, J.B. Caussin, E.~Lauga, D.~Bartolo, Phys. Rev. Lett. \textbf{110},
  038101 (2013)

\bibitem{2013Giomi_PRL}
L.~Giomi, M.J. Bowick, X.~Ma, M.C. Marchetti, Phys. Rev. Lett. \textbf{110},
  228101 (2013)

\bibitem{2014Zoettl_PRL}
A.~Z\"ottl, H.~Stark, Phys. Rev. Lett. \textbf{112}, 118101 (2014)

\bibitem{2000WuLi}
X.L. Wu, A.~Libchaber, Phys. Rev. Lett. \textbf{84}, 3017 (2000)

\bibitem{2009LeEtAl_Gold}
K.C. Leptos, J.S. Guasto, J.P. Gollub, A.I. Pesci, R.E. Goldstein, Phys. Rev.
  Lett. \textbf{103}, 198103 (2009)

\bibitem{2010DuPuZaYe}
J.~Dunkel, V.B. Putz, I.M. Zaid, J.M. Yeomans, Soft Matter \textbf{6}, 4268
  (2010)

\bibitem{2011ZaDuYe}
I.M. Zaid, J.~Dunkel, J.M. Yeomans, J. R. Soc. Interface \textbf{8}, 1314
  (2011)

\bibitem{2009SoAr}
A.~Sokolov, I.S. Aranson, Phys. Rev. Lett. \textbf{103}(14), 148101 (2009)

\bibitem{2009Rafai}
S.~Rafai, L.~Jibuti, P.~Peyla, Phys. Rev. Lett. \textbf{104}, 098102 (2010)

\bibitem{2011Aranson_PRE}
S.D. Ryan, B.M. Haines, L.~Beryland, F.~Ziebert, I.S. Aranson, Phys. Rev. E
  \textbf{83}(5), 050904(R) (2011)

\bibitem{2012Foffano_EPJE}
G.~Foffano, J.S. Lintuvuori, A.N.M. nd~K.~Stratford, M.E. Cates, D.~Marenduzzo,
  Eur. Phys. J. E \textbf{35}, 98 (2012)

\bibitem{2013Wioland_PRL}
H.~Wioland, F.G. Woodhouse, J.~Dunkel, J.O. Kessler, R.E. Goldstein, Phys. Rev.
  Lett. \textbf{110}, 268102 (2013)

\bibitem{2013Thampi_PRL}
S.P. Thampi, R.~Golestanian, J.M. Yeomans, Phys. Rev. Lett. \textbf{111},
  118101 (2013)

\bibitem{2014Lushi_PNAS}
E.~Lushi, H.~Wioland, R.E. Goldstein, Proc. Natl. Acad. Sci. USA
  \textbf{111}(27), 9733 (2014)

\bibitem{2013Dunkel_NJP}
J.~Dunkel, S.~Heidenreich, M.~B\"ar, R.E. Goldstein, New J. Phys. \textbf{15},
  045016 (2013)

\bibitem{Landau59}
L.D. Landau, E.M. Lifshitz, \emph{Theory of Elasticity} (Pergamon, London,
  1959)

\bibitem{2006Aranson}
I.S. Aranson, L.S. Tsimring, Rev. Mod. Phys. \textbf{78}, 641 (2006)

\bibitem{1977SwiftHohenberg}
J.~Swift, P.C. Hohenberg, Phys. Rev. A \textbf{15}(1), 319 (1977)

\bibitem{1995TonerTu_PRL}
J.~Toner, Y.~Tu, Phys. Rev. Lett. \textbf{75}(23), 4326 (1995)

\bibitem{1998TonerTu_PRE}
J.~Toner, Y.~Tu, Phys. Rev. E \textbf{58}(4), 4828 (1998)

\bibitem{2005ToTuRa}
J.~Toner, Y.~Tu, S.~Ramaswamy, Ann. Phys. \textbf{318}, 170 (2005)

\bibitem{2010Ramaswamy}
S.~Ramaswamy, Annu. Rev. Cond. Mat. Phys. \textbf{1}, 323 (2010)

\bibitem{2014Grossmann_PRL}
R.~Gro\ss{}mann, P.~Romanczuk, M.~B\"ar, L.~Schimansky-Geier, Phys. Rev. Lett.
  \textbf{113}, 258104 (2014)

\bibitem{2002Ra}
R.A. Simha, S.~Ramaswamy, Phys. Rev. Lett. \textbf{89}(5), 058101 (2002)

\bibitem{2013Henann_PNAS}
D.L. Henann, K.~Kamrin, Proceedings of the National Academy of Sciences
  \textbf{110}(17), 6730 (2013)

\bibitem{1996Risken}
H.~Risken, \emph{The Fokker-Planck Equation: Methods of Solutions and
  Applications}, 2nd~edn. (Springer, Berlin, 1996)

\bibitem{2004Frisch}
U.~Frisch, \emph{Turbulence} (Cambridge University Press, Cambridge, England,
  2004)

\bibitem{Abraham1988}
R.~Abraham, J.~Marsden, T.~Ratiu, \emph{Manifolds, Tensor Analysis, and
  Applications}, Vol.~75 of \emph{Applied Mathematical Sciences} (Springer,
  1989)

\bibitem{Canuto1987}
C.~Canuto, M.~Hussaini, A.~Quarteroni, T.~Zang, \emph{Spectral Methods in Fluid
  Dynamics} (Springer-Verlag, Berlin, 1987)

\end{thebibliography}
\end{document}